\documentclass[conference]{IEEEtran}
\IEEEoverridecommandlockouts
\usepackage{cite}
\usepackage{amsmath,amssymb,amsfonts}
\usepackage{algorithmic}
\usepackage{graphicx}
\usepackage{orcidlink}
\usepackage{textcomp}
\usepackage{xcolor}
\usepackage{subcaption}
\usepackage{multirow}
\usepackage{array}  
\usepackage{cprotect}

\def\BibTeX{{\rm B\kern-.05em{\sc i\kern-.025em b}\kern-.08em
    T\kern-.1667em\lower.7ex\hbox{E}\kern-.125emX}}

\begin{document}

\title{Integration of quantum random number generators with post-quantum cryptography algorithms\\

\thanks{This work was partially funded by CEX2019-000910-S (MCIN/AEI/ 10.13039/501100011033), Fundació Cellex, Fundació Mir-Puig, and Generalitat de Catalunya through CERCA; Departament de Recerca i Universitats de la Generalitat de Catalunya (2021 SGR 01458); European Union (QSNP, 101114043). This study was supported by MCIN with funding from European Union NextGenerationEU (PRTR-C17.I1) and by Generalitat de Catalunya. Ayuda PRE2022-102373 financiada por MCIN/AEI/ 10.13039/501100011033 y por el FSE+.}
}

\author{
    Paula Alonso Blanco\IEEEauthorrefmark{1}\orcidlink{0000-0002-2301-1517}, 
    Luis Trigo Vidarte\IEEEauthorrefmark{1}\orcidlink{0000-0003-3686-3820}, 
    Marc Romeu Casas\IEEEauthorrefmark{2}, 
    José Ramón Martínez Saavedra\IEEEauthorrefmark{2}\orcidlink{0000-0001-7966-9038},\\
    Fernando de la Iglesia\IEEEauthorrefmark{2}\orcidlink{0000-0002-6127-9493},
    Jordi Mur-Petit\IEEEauthorrefmark{3}\orcidlink{0000-0002-4018-1323},
    Valerio Pruneri\IEEEauthorrefmark{1}\IEEEauthorrefmark{4} \orcidlink{0000-0002-6425-9332},\\
    \IEEEauthorrefmark{1}ICFO - The Institute of Photonic Sciences, 08860 Castelldefels, Barcelona, Spain \\
    \IEEEauthorrefmark{2}Quside Technologies SL, Mediterranean Technology Park, 08860 Castelldefels, Barcelona, Spain\\
    \IEEEauthorrefmark{3}Nestlé IT Innovation and Enterprise Architecture, 08950 Esplugues de Llobregat, Barcelona, Spain\\
    \IEEEauthorrefmark{4}ICREA - Catalan Institution for Research and Advanced Studies, Barcelona, Spain \\
}

\IEEEoverridecommandlockouts
\IEEEpubid{\makebox[\columnwidth]{979-8-3315-9777-1/25/\$31.00 \copyright2025 European Union \hfill} \hspace{\columnsep}\makebox[\columnwidth]{ }}

\maketitle
\IEEEpubidadjcol

\begin{abstract}
As quantum technologies advance, the security of popular cryptographic protocols becomes more threatened by the capabilities of Cryptographically Relevant Quantum Computers (CRQCs). In this scenario, Post-Quantum Cryptography (PQC) has become a potential solution to prolong the life of existing Public Key Infrastructure (PKI) systems. However, PQC protocols depend on high-quality randomness for key generation and encapsulation procedures, with the quality of the entropy source potentially having a profound impact on the security of the overall system. In this work, we demonstrate a proof-of-concept enabling the incorporation of Quantum Random Number Generation (QRNG) devices within communication networks using PQC-based Transport Layer Security (TLS).Using open-source cryptographic libraries and commercial QRNG hardware, we demonstrate their use as entropy sources via an Entropy-as-a-Service (EaaS) model. We highlight two particular use cases: a fully virtualized private PKI network and a connection to an external PQC-enabled server. Experimental results show that EaaS QRNG enables real-time entropy monitoring and quality assessment in cryptographic management systems, with negligible impact on TLS handshake time.
\end{abstract}


\section{Introduction}
\IEEEpubidadjcol
Ensuring secure interactions between users of a scalable communication network is far from a trivial task. Public Key Infrastructure (PKI) systems propose a framework that relies on asymmetric cryptographic algorithms during part of the exchanges and on trusted Certificate Authority (CA) entities. At the cost of expanding the environment of trust, the use of CAs circumvents the problem of distributing preshared keys between users, making PKIs a flexible solution for large-scale networks such as the Internet, accommodating the security needs of a wide range of users. On the other hand, reliance on asymmetric cryptography makes interactions vulnerable to specific computational cryptoanalysis attacks, in particular \textit{harvest now, decrypt later} attacks \cite{Mosca2015} are already a risk. Shor's algorithm \cite{Shor1997} running on a cryptographically relevant quantum computer (CRQC) can solve factorization and discrete logarithm problems efficiently, concepts that lay at the heart of the security assumptions in the most commonly used cryptographic protocols today, such as the Rivest-Shamir-Adleman (RSA) and elliptic curves (EC) family suites. 

A strategy to mitigate this risk in PKI systems is migrating to new cryptographic algorithms that depend on mathematical problems expected to be significantly harder to solve than their predecessors, even in the presence of CRQCs. The new algorithms are typically encompassed under the terms Post-Quantum Cryptography (PQC) and Quantum-Safe Cryptography (QSC). The efforts are coordinated by different institutions, such as NIST in the US, to create an ecosystem of protocols and solutions that evolve in a crypto-agile fashion, adapting to changes. These protocols can be combined with the previous standards in a hybrid manner to provide a fallback solution or support legacy devices. The main targets so far are providing Key Encapsulation Mechanisms (KEM) to distribute symmetric keys between users and Digital Signature Algorithms (DSA) to offer authentication and integrity features. Symmetric cryptography algorithms such as the Advanced Encryption Standard (AES) are not currently considered vulnerable.

\subsection{Randomness in cryptography}

Although the different PQC proposals might differ in their mathematical foundations, they must produce bit strings as unpredictable as possible in their initial steps. In order to generate these random sequences, they rely on entropy sources of sufficient quality for cryptographic applications. Previous studies have shown that low-entropy sources can facilitate the cryptanalysis of specific PQC algorithms \cite{cryptoeprint:2024/1229}. 

Among the different entropy source solutions, quantum random number generation (QRNG) devices offer two main advantages over other alternatives: (1) they produce randomness based on intrinsically random phenomena of quantum nature, and (2) they can be reliably calibrated, providing unpredictability indicators beyond statistical tests. 

Quantum entropy generated by reliable devices allows us to build trust based solely on the assumption that certain quantum phenomena are intrinsically random. For high-performance applications, confidence in the fabrication process of the QRNG and a solid understanding of its entropy source help interpret its output and strengthen trust in the randomness it provides.

State-of-the-art QRNG devices provide unpredictability indicators that complement statistical tests and enable real-time monitoring, ensuring the generated randomness is above a set threshold. Existing commercial devices can already provide random bits at high rates and competitive costs, but they need to probe some physical phenomena, making them require hardware components. This contrasts with the nature of PKI systems, which could be implemented entirely in software, although hardware accelerators are common. 

QRNG providers can offer devices suitable to many environments, but it is necessary to interact efficiently and safely with the applications where randomness will be used. To facilitate QRNG deployment, initiatives like QRNG Open API \cite{QRNG-OPENAPIweb} have emerged. Apart from integrating QRNG devices in the final products, using entropy sources servers is also considered, leading to the concept of entropy as a service (EaaS) \cite{NIST_EaaS}.

\subsection{Scope of this paper}
The primary motivation of this work is to serve as a proof-of-concept for users interested in integrating monitored QRNG devices in their networks, demonstrating that they can be seamlessly integrated into a network and provide quantum entropy to PQC algorithms, offering higher security standards. We describe how to do this integration using open-source tools, present two relevant use cases that showcase the use of EaaS QRNG randomness in PQC protocols, and present experimental results. We do not intend to benchmark PQC implementations \cite{Benchmarking_PQC_TLS}, detailed descriptions of the use of open-source tools and QRNG devices \cite{10.1007/978-3-030-44223-1_5, SPACE:GonzalezWiggers22, QRNG_TLS_UC3M} or to generalize our study to cover all possible scenarios of QRNG integration in PKI networks. 

\section{Methodology}
The generation of a shared secret key to use symmetric cryptography algorithms such as AES, can be accomplished using the Transport Layer Security (TLS) protocol. TLS follows a client-server model and aims to establish a shared random secret, the master key, from which session keys are derived. This secret is securely transmitted using a Key Encapsulation Mechanism (KEM) and relies on digital certificates and Digital Signature Algorithms (DSA). More recent proposals like KEMTLS \cite{CCS:SchSteWig20,SPACE:GonzalezWiggers22} can also use PQC KEMs instead of PQC DSA to improve efficiency.

Well-maintained open-source tools such as \verb|OpenSSL| \cite{openssl_web} provide most of the cryptographic suites accepted by TLS and support the implementation of clients, servers, and certificate management. PQC protocols were not natively supported by \verb|OpenSSL 3.2.0-beta1| at the time of the experiments, so we used the \verb|libOQS 0.9.0| library \cite{liboqs_github} to integrate pre-standardization versions of PQC algorithms. To integrate commercial QRNG devices in this setup, we adapted the \verb|libOQS| library to use the QRNG device as the source of random numbers for its algorithms.

\subsection{QRNG integration and use cases}
The experiments were performed using Quside Garnet\texttrademark\ PCIE 400, with a nominal rate of extracted random bits of 290 Mbps. This rate is far greater than the required in conventional scenarios, where algorithms typically demand fewer than 100 bytes per connection. For this reason, we decided to integrate the QRNG device in a server running an entropy source service in the line of EaaS approaches. We constructed virtual networks using Virtual Machines (VM) from this starting point to emulate relevant scenarios flexibly. Among the scenarios tested, we present two particularly relevant use cases.

\begin{figure}[t]
\centering
\includegraphics[width=\linewidth]{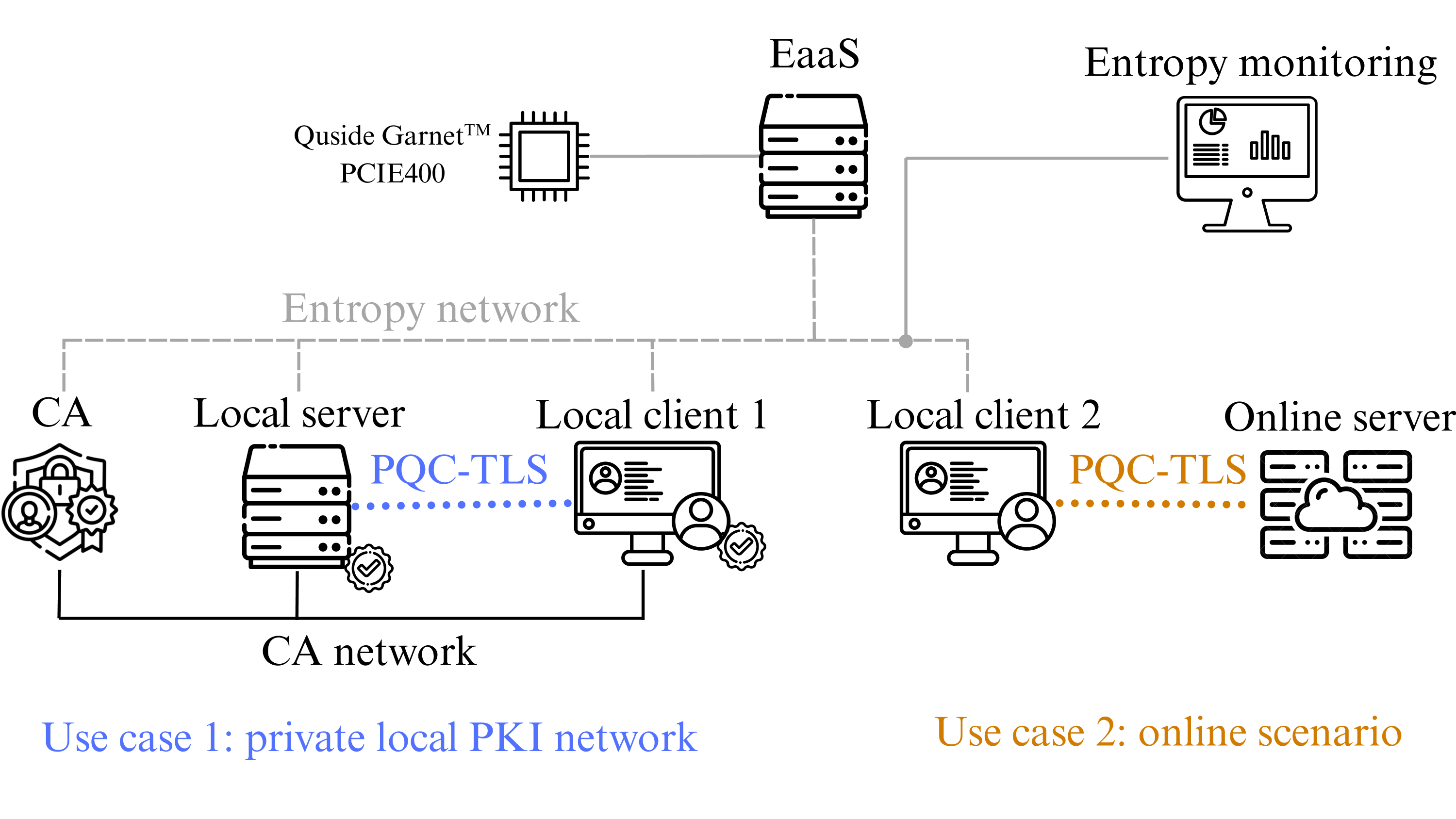}
\caption{Network architectures used in the two experiments. In the left-hand part, a local network with a self-signed CA distributes certificates to a local server and client, enabling a post-quantum cryptography TLS connection. The right-hand figure shows a remote client connecting to the online server. The entropy network provides Entropy as a Service (EaaS) and supports real-time monitoring in both scenarios.}

\label{fig:network_architectures}
\end{figure}
In use case 1, we consider a small-scale private PKI network that can represent the internal network of a company, where all users have access to the random numbers of the entropy source when they are using PQC algorithms. On the left side of Fig.~\ref{fig:network_architectures}, we show the scheme used for this scenario in blue. 
We observe the PKI structure where a VM server acts as a self-signed CA to later provide the necessary PQC certificates to the CA server, the local server, and the local client 1 through the CA network. After all entities are certified, a PQC-TLS connection is established between Local Server and Local Client 1 using different KEMs. 

The experiment is done by starting several processes in various server ports, each using a different certificate and KEM configuration for the TLS connection. This way, we could test several PQC-TLS connections between server and client to obtain connection statistics for different parameters.

In use case 2, we test the interaction between a local client using a quantum entropy source and an external server supporting PQC protocols. The open-source project \verb|OQS| \cite{stebila2017oqs} offers an online server that supports most proposed PQC algorithms and their hybrid combinations. This server provides the CA certificate to enable authentication using PQC digital signatures. On the right side of Fig.~\ref{fig:network_architectures}, highlighted in light orange, we illustrate the modified local network setup in which the client connects to the \verb|OQS| online server using PQC KEMS. This setup enables a realistic evaluation of bandwidth and processing overhead when using PQC and hybrid algorithms with external entities.

Finally, Fig.~\ref{fig:network_architectures} highlights, in grey, a node that serves as an entropy source (EaaS) through the entropy network in both use cases, delivering QRNG-based randomness to all PQC algorithms and enabling real-time monitoring during PQC-TLS testing.

\subsection{Metrics and algorithms}
Using the open-source tool \verb|Wireshark|, we studied the network traffic in both experiments and examined several parameters related to the performance of the PQC algorithms, including the handshake duration, the bandwidth, and the number of packets exchanged during the communication. Moreover, we wanted to describe how the EaaS impacts the total TLS handshake time. To do so, we measured the latency of the entropy provided by the QRNG and divided it by the time to perform the handshakes to obtain the latency overhead of this service. We define the handshake time as the complete duration of the TLS communication from the first packet sent to the key establishment. Finally, we analyzed the randomness demand placed on the QRNG during the encapsulation step, a critical phase of PQC-based TLS in which the client generates the shared secret using fresh entropy. These parameters were evaluated across the PQC-KEMs used. We also analyzed the contribution of the QRNG device to the overall EaaS latency in order to isolate device performance from network-induced delays.

An open question in the deployment of PQC is the timeline and process for deprecating so-called traditional cryptography algorithms (like RSA or ECC), which benefit from decades of crypto analysis. One approach proposes moving directly to PQC, while a more conservative approach in terms of cybersecurity is the so-called hybrid approach, which combines traditional algorithms with PQC ones. 
In our experiments, we chose to use the hybrid approach, selecting only a subset of all the available PQC algorithms for simplicity. These algorithms are categorized into different Security Levels (SL), which allows for a consistent comparison of the security guarantees provided by each algorithm implementation. This classification defines five security levels, ranging from Level 1 to Level 5. This work focuses on SL 3 for the PQC algorithms, which represents a moderately conservative choice within the standardization spectrum.
\section{Results}
\begin{figure}[t]
\centerline{\includegraphics[width=\linewidth]{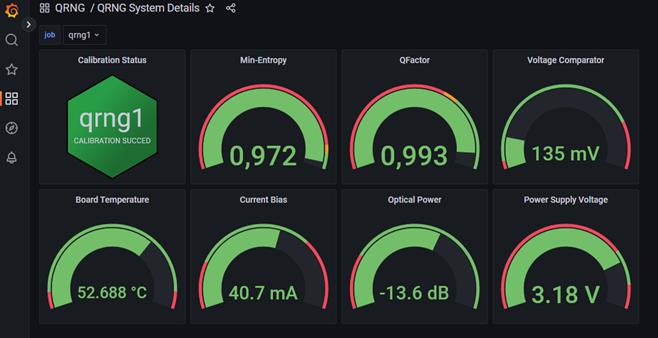}}
\caption{Grafana dashboard with QRNG metrics. Quantum min-entropy is the physical measure of the unpredictability of the provided random bytes. The Q-Factor is a statistical measure of randomness. Left-side metrics correspond to photonic components of the QRNG.}
\label{fig:qrng-dashboard}
\end{figure}

In our experiments, the monitoring of the QRNG ensured that key parameters never dropped below critical thresholds. The device can provide the quantum min-entropy of the served random bytes, which is a measure of their unpredictability. The QRNG used has an average quantum min-entropy of 0.93 bits per bit. In addition, each request for random bytes served by the QRNG was provided with statistical metrics of the generated randomness. Other physical metrics related with the quantum system were monitored during the experiments. Fig. \ref{fig:qrng-dashboard} shows a dashboard with the monitored metrics.

\begin{table}[t]
\caption{Randomness demand for encapsulation per KEM, as well as the temporal and QRNG overhead from EaaS in TLS connections. }
\begin{center}
\begin{tabular}{|c|c|c|c|}
\hline
\textbf{} & \textbf{Encapsulation}  & \textbf{EaaS } & \textbf{QRNG temporal} \\
\textbf{Algorithm} & \textbf{random bytes}  & \textbf{ latency (ms)} & \textbf{ overhead $(10^{-6})$} \\
\hline
kyber768 &32 &30.53 & 4.19 \\
bikel3 & 64 & 40.09 & 6.39 \\
hqc192 & 24 & 19.61 & 9.79 \\
frodo976aes & 24 & 19.61 & 9.79 \\
frodo976shake & 24 & 19.61 & 9.79 \\
\hline
\end{tabular}
\label{tab: latency pqc}
\end{center}
\end{table}

\subsection{EaaS overhead in TLS}
Table \ref{tab: latency pqc} shows the number of random bytes required by each KEM in the \verb|libOQS| library to generate the encapsulation function and the average time to generate them. The generation time consistently remains at or below $41$ ms, with most latency attributed to network delays. The QRNG overhead accounts for, at most, a fraction of only $9.79\cdot10^{-6}$ of the total latency. This result highlights that the QRNG performance outperforms the needs for this class of application where most of the latency comes from the network. 

\begin{figure}
    \centering
    \includegraphics[width=\linewidth]{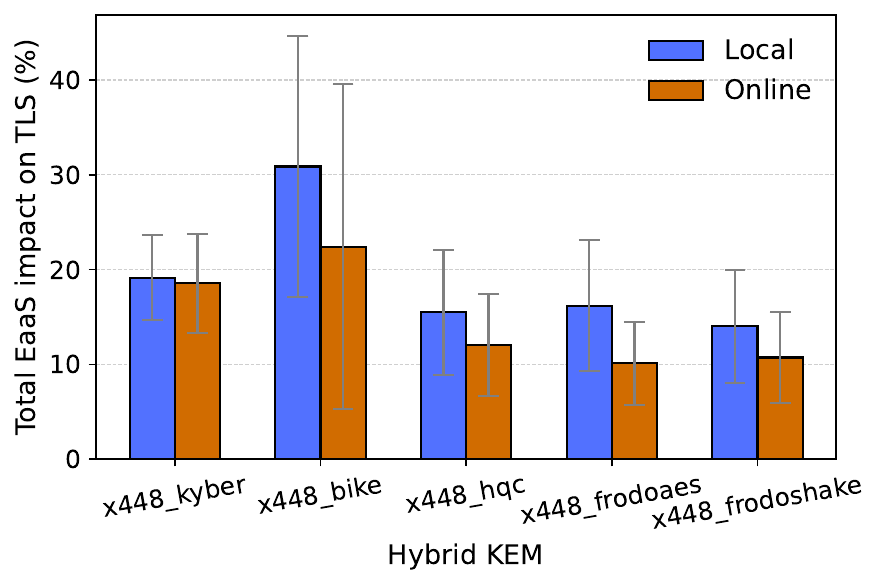}
    \cprotect\caption{Overhead of using Entropy as a Service (EaaS) in TLS connections with different hybrid KEMs, evaluated in both a local virtual network and an online server setup. Bars show the ratio between EaaS latency (generation + transmission) and the total TLS handshake time. The server uses \verb|p384-dilithium3| for signing. All KEMs, both traditional and post-quantum, are at Security Level 3. Error bars show ratio uncertainty via standard error propagation.}
    \label{fig:overhead eaas}
\end{figure}

Figure \ref{fig:overhead eaas} shows the contribution of entropy distribution to the total TLS handshake time in both use cases. In the local private PKI scenario, the overhead remains consistently below $32\%$, dropping below $15\%$ in cases such as \verb|frodoshake|. The online setting exhibits a similar trend with lower relative overhead, reaching even $10\%$. This difference is due to higher overall handshake latency caused by increased client–server distance, which reduces the relative impact of the entropy network.

Finally, we observe comparable trends in both local and online environments, with the expected increase in total handshake time in the online case due to network-induced delays inducing a minor impact of EaaS on TLS.

\subsection{Effect of using PQC in TLS}
\begin{figure}[t]
\centerline{\includegraphics[width=\linewidth]{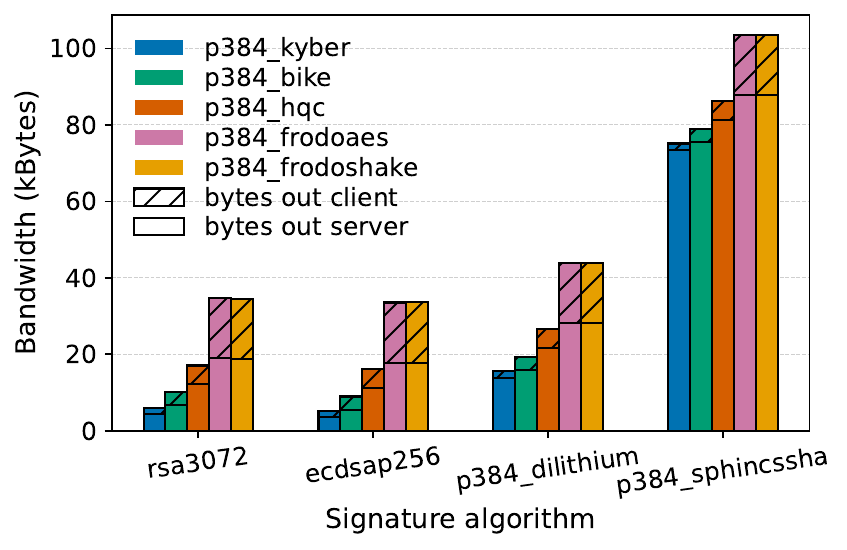}}
\caption{Bandwidth, in kilobytes, used to establish a TLS connection with the online server for various KEM and DSA combinations. Hybrid KEM and DSA use security level (SL) 3, while traditional signatures, RSA and ECDSA, use SL 1. Dashed bars indicate client-to-server bytes; solid bars show server-to-client bytes. The total bar height represents the full communication bandwidth.}
\label{fig: bandwidth online}
\end{figure}

Figure \ref{fig: bandwidth online} illustrates the number of bytes exchanged during the TLS handshake. Each bar is divided into two segments: the upper dashed portion corresponds to the bytes sent from client to server, while the lower solid portion represents the bytes sent from server to client. The figure compares multiple hybrid KEM configurations grouped by the DSA used, both classical (RSA and ECDSA) and PQC (Dilithium and SPHINCS+). All KEMs employed are hybrid PQC algorithms.

Digital certificates significantly influence overall bandwidth consumption, mainly when using hybrid approaches. For instance, in the case of digital signatures, the use of Dilithium increases the transmitted data by over 10 kB (kilobytes) compared to RSA or ECDSA and Sphincs by more than 70 kB. Among the PQC KEMs, Kyber has the lowest bandwidth contribution, followed by BIKE and HQC, with both Frodo variants introducing a significantly higher overhead, showing a relative increase of nearly 28 kB compared to Kyber. These increases are primarily due to the size of public keys and signatures in PQC schemes.

We compare the network impact of using PQC digital signatures such as Dilithium and SPHINCS+ against traditional signatures like RSA and ECDSA-256, highlighting the expected rise in bandwidth requirements. While our results focus on SL 3, we note that selecting higher SLs increases the size of the cryptographic material, thereby increasing bandwidth. Besides impacting bandwidth, this increase also results in a greater number of packets exchanged. As larger handshake messages exceed standard packet size limits, they must be fragmented, resulting in a more complex and voluminous message exchange during connection establishment.

Several previous studies have provided a more in-depth analysis of the performance of PQC algorithms \cite{10.1007/978-3-030-44223-1_5, leise2024entropy}.

\section{Conclusions}
In this work we have presented a proof-of-concept for the integration of QRNG devices with PQC systems run over open-source TLS implementations. We have demonstrated the consistent distribution of entropy over the network through an EaaS framework that PQC and hybrid algorithms can directly use without significant changes to existing software infrastructures. Our approach relies on well-maintained software such as \verb|OpenSSL| and \verb|libOQS|, which we adapted to enable QRNGs as primary entropy sources.

We have outlined two pertinent use cases: a local PKI virtual network and an external PQC server connection. In both cases, we have examined key performance indicators like TLS handshake latency, bandwidth consumption, packet number, entropy response time, and randomness overhead over the TLS handshake latency. Our results show that the EaaS approach introduces temporal overheads of typically less than 30\% in our scenarios, of which a negligible fraction $<10^{-5}$ corresponds to the QRNG, demonstrating that QRNG devices can be efficiently combined with PQC algorithms to improve the trustworthiness of these schemes. 

\section*{Disclosures}
VP is shareholder and board member of Quside Technologies SL.

\bibliographystyle{IEEEtran}
\bibliography{paper.bib}

\end{document}